**Improving the Accessibility of Dating Websites for Individuals with Visual Impairments**

*Gyanendra Shrestha, Soumya Tejaswi Vadlamani*
*Department of Computer Science, Florida State University, Florida 32306, USA*

## 1. Introduction

People now frequently meet and develop relationships through online dating. Yet, due to their limited accessibility, utilizing dating services can be difficult and irritating for people with visual impairments. The significance of the research issue can be attributed to the fact that dating websites are becoming more and more common and have a significant impact on how people establish romantic connections. It can be challenging for people with visual impairments to use dating services and develop lasting relationships because many of them are not created with their requirements in mind. We can encourage people with visual impairments to participate more completely in online dating and possibly enhance the success of their romantic relationships by making dating websites more accessible. There is some existing implementation that can automatically recognize the facial expression, age, gender, presence of child(ren) and other common objects from a profile photo in a dating platform. The goal of this project is incorporate additional features (presence of any common pets, indoor vs. outdoor image) to further enhance the capability of existing system and come up with test viable solutions to accessibility issues that people with visual impairments face when using dating websites.

## 2. Literature Review

Moore et al. (2019) [1] identified the problem of how Online dating platforms is a barrier for the visually impaired people as it relies heavily on visual data. So, they proposed a framework which can be used by such individual to meet new people. They created an application which works via screen shot that is converted to a text and then to speech through which an informed decision can be made by the user.

Abraham et al. (2020) [2] proposed a framework for speech-based feedback system to enable the blind community to experience the world on their own. This paper recognized walkable spaces, texts, converts text to speech, and locates particular kinds of objects, among other things. The system contains a Raspberry Pi, camera module and audio jack, that will be worn by the visually impaired user.

In Chaganti et al. (2020) [3] image classification is done using Support Vector Machine (SVM) and Convolutional Neural Network (CNN), results were compared and evaluated. They find that Neural Networks and CNN reveal much more potential than SVM. Even though SVM is a very strong technique among traditional machine learning algorithms, the accuracy of its use in image classification still falls behind, compared to deep learning techniques.

Bhardwaj, Deepika et al. (2021) [4] proposed a new approach for the classification of the indoor-outdoor images into their classes using VGG-16 CNN Model. The classification of scenes is done using computer vision as a stage and then the machine uses deep learning to describe the scene accordingly.

## 3. Methodology

In this project we used a VGG16 deep CNN model [5] for image classification problems to identify any presence of pet animals and indoor-outdoor scenes from user profile images taken from dating websites. To build our learning networks, we used the pre-trained VGG16 convolutional neural network.

### 3.1. Data Preprocessing

Data preprocessing is a required first step in any machine learning approach which involves steps like labeling images, resizing them, converting them to NumPy arrays for input into a deep learning model.

### 3.2. Data Augmentation

To expand the diversity of our training dataset, we applied common data augmentation techniques [6] such as reflection, rotation, color adjustment, and brightness modification. These methods are often used to introduce variability and increase the size of the training dataset. By performing these operations, we were able to generate additional instances of each image that were sufficiently distinct from the original, which ultimately improved the accuracy of our model by enabling it to generalize better to new and unseen images.

### 3.3. Convolutional Neural Network (CNN)

CNN is one of the most popular supervised deep learning algorithms for image classification. It consists of multiple convolutional layers and pooling layers, following full connected networks as hidden layers and the output layer. The convolution and pooling layers extract/learn features and fully connected layers learn to make predictions. The parameters of CNN like the number of layers, number of neurons per layer, max pooling window size, filter size in convolution layer etc. can be adjusted during the experiment to boost the detection effect of the classifier accordingly.

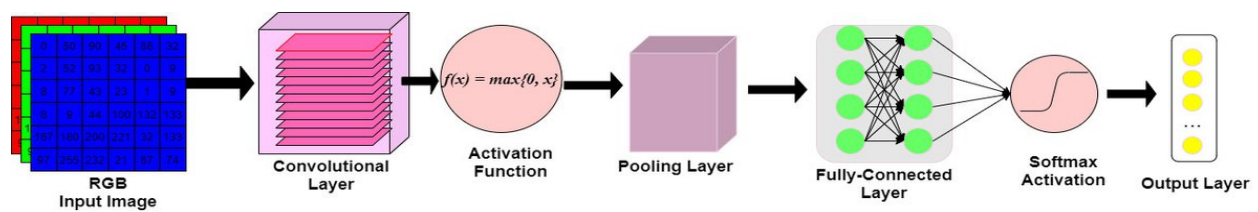

Figure 1: General CNN Architecture

### 3.4. Neural network model VGG16

VGG16 is a CNN 16-layer, pre-trained 100-class model. It was proposed by Karen Simonyan and Andrew Zisserman from Oxford University "Very Deep Convolutional Networks of LargeScale

Image Recognition". The 16 in VGG16 refers to 16 layers that have weights. This model achieves 92.7% top-5 test accuracy on the ImageNet dataset which contains 14 million images belonging to 1000 classes. It takes input tensor size as 224, 244 with 3 RGB channels.

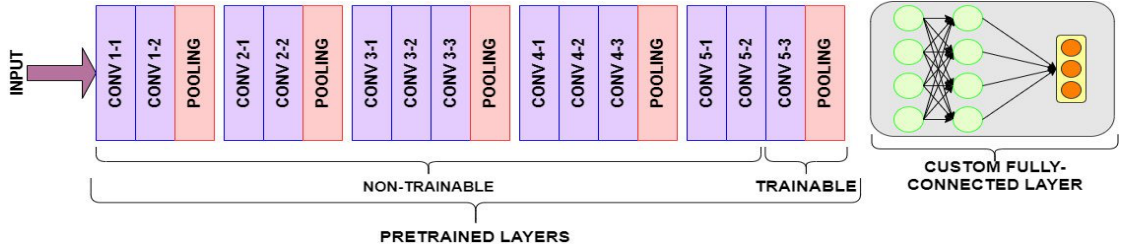

**Figure 2: VGG16 model**

### 3.5. Loss Function

We used binary cross-entropy loss function for training our proposed CNN model. The loss formula is defined as:

$$\text{Loss} = -\frac{1}{N} \sum_{i=1}^{N} \left( y_i * \log\left((P(y_i))\right) + (1 - y_i) * \log\left(1 - P(y_i)\right) \right) \quad (1)$$

Where $N$ is the number of samples in the dataset, $y$ is the true label (either 0 or 1), and $P(y)$ is the predicted probability of the positive class (usually represented by 1). The first term in equation 1 calculates the loss for positive class and the second term calculates the loss for the negative class.

### 3.6. Text-To-Speech

Since our goal is to assist visually impaired people by retrieving information from photos the prediction result should be in the form easily accessible to them which obviously is sound. gTTS (Google Text-To-Speech) is used to convert textual contents to audio, which is the final output to the visually impaired user.

## 4. Implementation

The deep learning model we have developed follows a two-phase approach, namely the training phase and prediction phase (as shown in Figure 3). In the training phase, the labeled images are fed into the deep learning algorithm to learn relevant features, resulting in a trained classifier. In the prediction phase, the trained classifier is used to make predictions on unseen input images to determine their corresponding labels.

Initially, we had designed our deep learning model to perform multilabel image classification for predicting both pet animals and indoor-outdoor scenes. However, this approach failed to accurately predict the presence of pet animals in our test images. Therefore, we decided to develop separate models for pet and scene classification, as shown in Figure 4. The idea was to first pass the images through the indoor-outdoor classifier to predict the scene in the image and then through the pet-animal classifier to predict the presence of any pet animals. However, the output of the pet-animal classifier did not meet our expectations, as its performance was poor. Our pet-classifier had 10 different classes, with uneven numbers of training images in each class. To improve the classifier's performance, we decided to focus solely on cats and dogs and renamed the model as the cat-dog classifier. The idea of using only cats and dogs for the pet-animal classifier is supported by the fact that they are the most common pet animals.

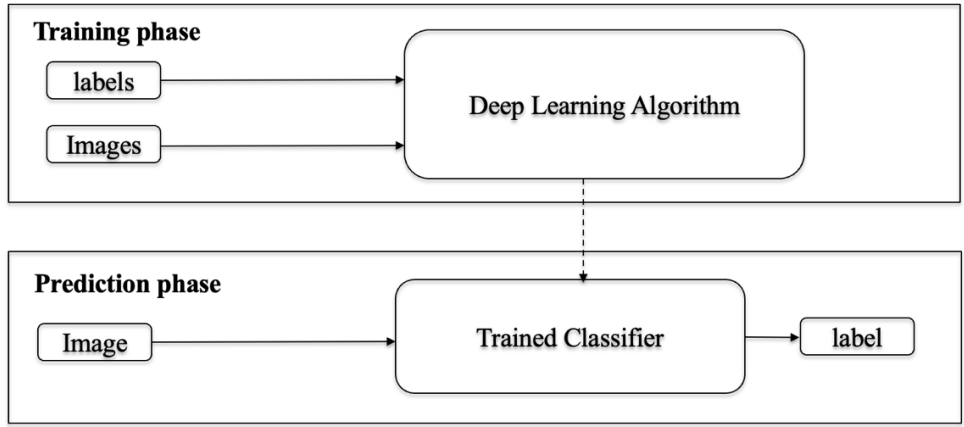

Figure 3: System Architecture

For fine-tuning the pre-trained VGG16 network, we removed the last layer of the network (Figure 2) and replaced them with a new classification layer that is specific to the problem of either cat-dog or indoor-outdoor classification. This new classification layer was trained using our training datasets.

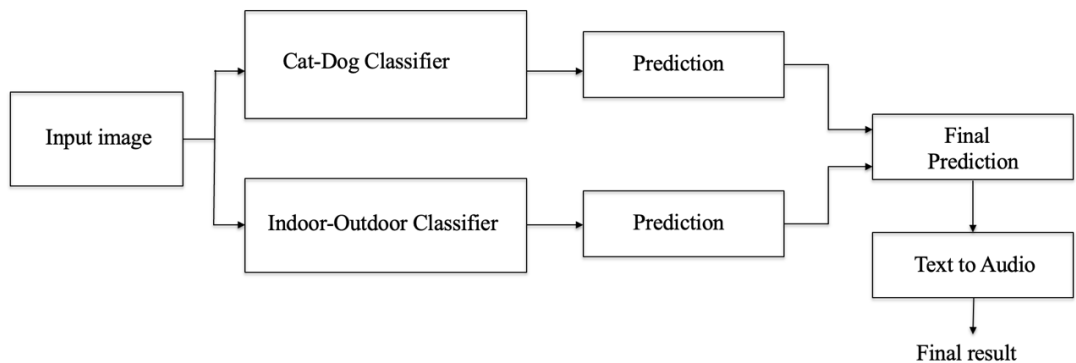

Figure 4: System Pipeline

**Algorithm:**

Step 1 Collect input images for training, validation. and testing
Step 2 Data Preprocessing and augmentation
Step 3 Implementing CNN model using transfer learning
Step 4 Test the images
Step 5 Predict final result in the form of audio output

## 5. Experiment

To run the training process and evaluation, we used a Google Compute Engine backend with GPU support. This system has 35.5 GB of system RAM and 15.0 GB of GPU RAM, which is sufficient for deep learning tasks involving large datasets and complex models. By using a GPU, we were able to speed up the training process and achieve results in a shorter amount of time compared to running the same code on a CPU.

The training parameters used in our experiment are as follows:
- Batch size: 32
- Epoch: 20
- Pre-trained weights: ImageNet
- Optimizer: Adam
- Learning rate: 0.0001

### 5.1. Datasets

To train a classifier that can distinguish between cats and dogs and indoor and outdoor scenes, we obtained the datasets from sources such as Kaggle and Google search. We collected images from the Animals-10 dataset [7], MIT Indoor Scenes [8] and Places365 datasets [9]. Animals-10 contains about 28K animal images belonging to 10 categories: dog, cat, horse, spider, butterfly, chicken, sheep, cow, squirrel, elephant. Image count for each category varies from 2K to 5 K units. MIT Indoor Scenes datasets contains 67 indoor categories, and a total of 15,620 images. The number of images varies across categories, but there are at least 100 images per category. Places365 datasets consist of 365 scene categories, where there are at most 5,000 images per category. For the cat-dog classifier, a total of 25,000 images (12,500 cats and 12,500 dogs) were used, while for the indoor-outdoor classifier, a total of 26,000 images (13,000 indoor and 13,000 outdoor) were used. An 80/20 split was used for training and validation, respectively. For testing purpose, we manually created a dataset downloading images from the google website.

### 5.2. Results and evaluation

Although we set the epoch value to 20 during our training experiment, both our cat-dog and indoor-outdoor models stopped training early after only 10 and 12 epochs, respectively. This was due to our training process being configured to stop early if the model's performance on validation set did not improve for certain number of epoch to prevent overfitting and save time on training. Figure 5 and Figure 6 illustrate the training statistics of the cat-dog and indoor-outdoor classifiers, respectively, which exhibit similar patterns. The validation accuracy was high even at the initial epoch due to the utilization of the VGG16 transfer learning network, which was originally trained on the ImageNet dataset. After epoch 2, the validation accuracy consistently remained lower than the training accuracy. Also, the validation loss was consistently higher than training loss indicating the model was not generalizing well to the validation data. As a result, the validation loss did not improve after epoch 8 for the cat-dog classifier and epoch 10 for the indoor-outdoor classifier, leading to early stopping to prevent overfitting. When our models stopped training, the precision and recall measures of the validation set were lower than those of the training set.

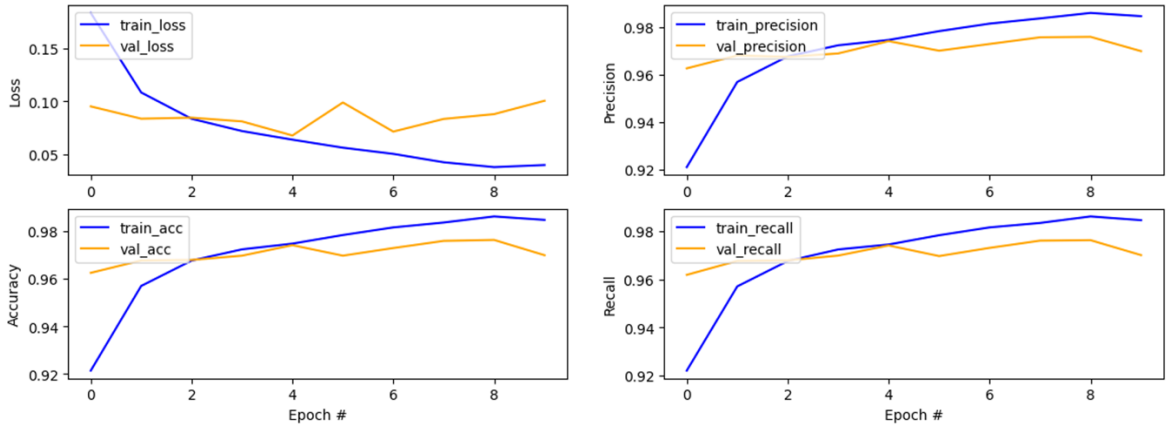

**Figure 5: Training Statistics of cat-dog classifier**

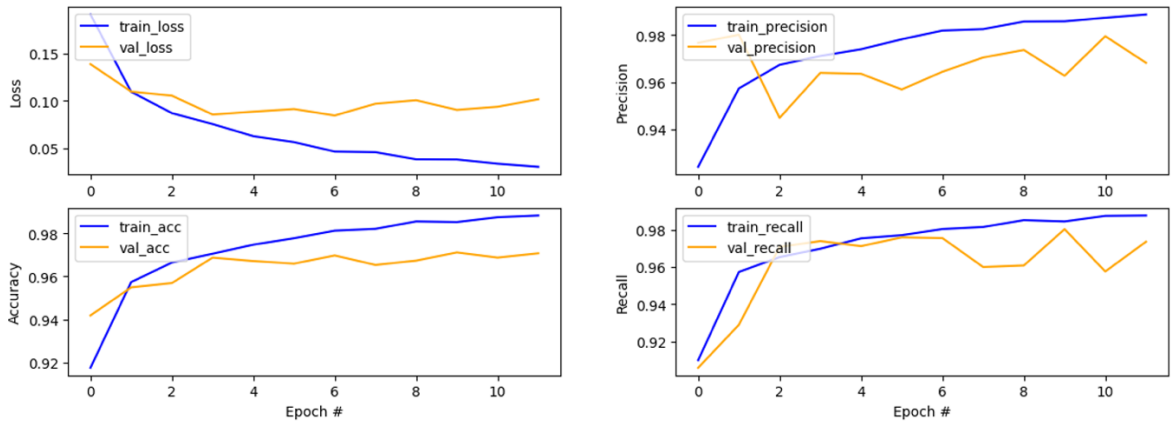

**Figure 6: Training Statistics of indoor-outdoor classifier**

The model's performance was evaluated using various metrics including accuracy, precision, recall, and F1-score.

**Table 1: Evaluation of cat-dog classifier on cat and dog images**

|          | precision | recall | f1-score | support |
|----------|-----------|--------|----------|---------|
| cat      | 0.99      | 0.98   | 0.99     | 1011    |
| dog      | 0.98      | 0.99   | 0.99     | 1012    |
| accuracy |           | **0.99** |        | 2023    |

Initially, we tested the performance of the cat-dog and indoor-outdoor classifiers separately. The cat-dog classifier was tested on a total of 2,023 images of cats and dogs, with only 29 images being misclassified, resulting in a very high accuracy score of 0.99. The evaluation summary can be found in Table 1. Similarly, we evaluated the performance of the indoor-outdoor classifier on a total of 2,001 images, with only 46 misclassifications resulting in an accuracy of 0.98. The evaluation summary is presented in Table 2.

**Table 2: Evaluation of indoor-outdoor classifier on indoor and outdoor images**

|  | precision | recall | f1-score | support |
|---|---|---|---|---|
| indoor | 0.97 | 0.98 | 0.98 | 1000 |
| outdoor | 0.98 | 0.97 | 0.98 | 1001 |
| accuracy |  | **0.98** |  | 2001 |

Finally, to evaluate the effectiveness of our combined classifier model, we assessed the performance of each individual classifier (cat-dog and indoor-outdoor classifier) for a given image. Our evaluation was based on a dataset of 200 realistic profile pictures of various individuals, equally divided among the four categories: cat-indoor, dog-indoor, cat-outdoor, and dog-outdoor. Table 3 displays the classification report for the cat-dog classes, which shows that the model has room for improvement. The precision for the cat class is 0.84, indicating that 84% of the instances classified as cats were actually cats. However, the recall for the cat class is 0.58, meaning that the model correctly identified only 58% of the actual cats. On the other hand, the precision and recall for the dog class are 0.68 and 0.89, respectively. This suggests that our model tends to miss lots of cat classes and misclassify them as dogs due to the equal number of images for both classes. The F1-score for the cat class is 0.69, and for the dog class, it is 0.77. The overall accuracy of the cat-dog classifier model is 0.73. Similarly, Table 4 presents the classification report for indoor-outdoor classes, which exhibits a similar pattern to the cat-dog classifier. Our indoor-outdoor classifier also misses many indoor classes to misclassify them as outdoor classes. However, the overall accuracy of indoor-outdoor classifier is 0.77, which is slightly better than that of the cat-dog classifier.

**Table 3: Evaluation of cat-dog classifier on profile images from dating website**

|  | precision | recall | f1-score | support |
|---|---|---|---|---|
| cat | 0.84 | 0.58 | 0.69 | 100 |
| dog | 0.68 | 0.89 | 0.77 | 100 |
| accuracy |  | **0.73** |  | 200 |

On closer examination of the images that were misclassified by our cat-dog and indoor-outdoor classifiers, we observed that they often had ambiguous properties (Figure 7 and Figure 8). For example, indoor images with plants and a very bright background were misclassified as outdoor, and similarly, images without a clear background were misclassified as indoor. The same was observed for cat and dog images where relevant features were hard to discern.

**Table 4: Evaluation of indoor-outdoor classifier on profile images from dating website**

|          | precision | recall | f1-score | support |
|----------|-----------|--------|----------|---------|
| indoor   | 0.95      | 0.54   | 0.69     | 100     |
| outdoor  | 0.70      | 0.97   | 0.81     | 100     |
| accuracy |           | **0.77** |        | 200     |

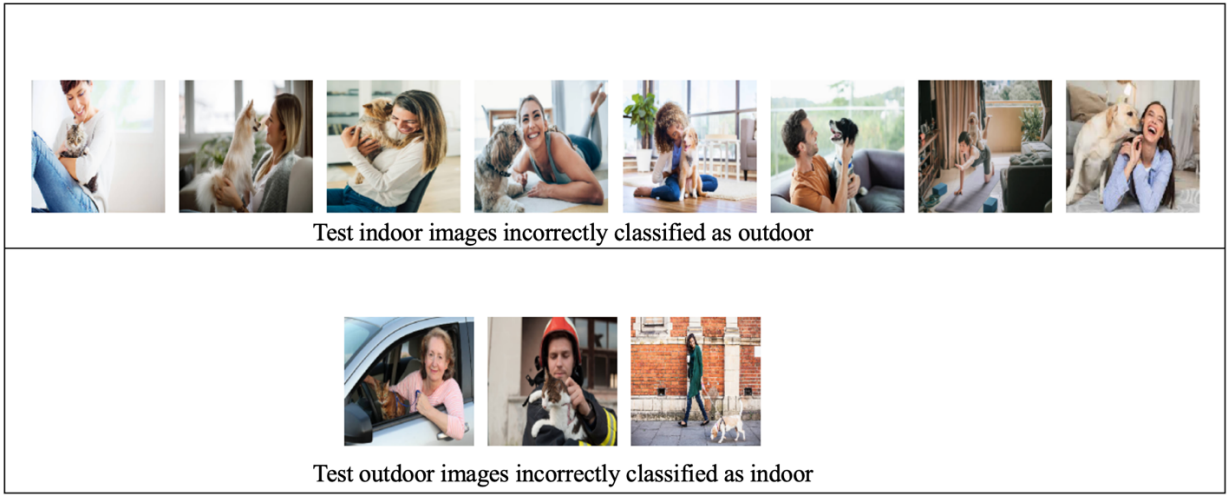

Figure 7: Indoor-outdoor test profile images misclassified by our model

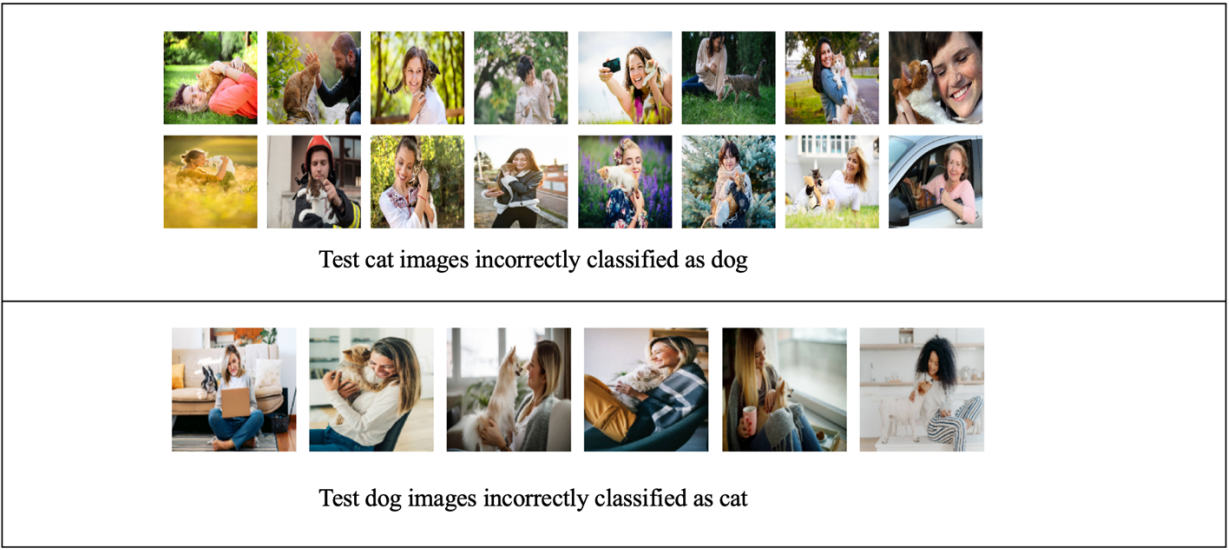

Figure 8: Cat-dog test profile images misclassified by our model

## 6. Conclusion

In this work, we presented a deep learning approach for labeling indoor and outdoor scenes and detecting the presence of cats or dogs in user profile images from Dating Websites. We extended previous image classification works by adding these new features and providing the labeling information as audio output to enhance the accessibility of Dating Websites for visually impaired individuals. Our model achieved a reasonable average accuracy rate of approximately 75%. However, the performance of our cat-dog classifier was found to be unsatisfactory on some negative test samples, with high probability values given even when there were no cats or dogs in the images. This might have been caused by overfitting, and further work is required to address this issue.

## 7. Future Work

Although VGG16 is a powerful architecture for image classification, exploring and combining different architectures like ResNet, DenseNet, or Inception could potentially improve the accuracy of the classification. Each architecture has its strengths and weaknesses and combining them can result in a more robust and accurate model. In addition to pet animals such as cats and dogs, there are many other popular classes of pet animals that could be included in the dataset. For example, birds, fish, rabbits, and reptiles are all commonly kept as pets. Including these additional classes could make the model more versatile and useful for a wider range of applications. Incorporating contextual information like object detection and semantic segmentation into VGG16 for scene classification could also improve the accuracy of the model. Object detection, for example, can help to identify and locate objects within an image, while semantic segmentation can provide information about the spatial relationships between objects. By incorporating this information into the classification process, the model can better understand the context in which the pet animals are located and improve its accuracy.

Overall, by exploring different architectures, incorporating additional classes of pet animals, and incorporating contextual information, we can improve the accuracy and robustness of the classification model for pet animal images. This could make the model more useful and applicable to a wider range of scenarios and applications.

# References

1. Moore, Meredith, et al. "The Blind Date: Improving the Accessibility of Mobile Dating Applications for Individuals with Visual Impairments." 2019 IEEE Global Conference on Signal and Information Processing (Global SIP). IEEE, 2019.
2. Abraham, Leo, et al. "VISION-wearable speech-based feedback system for the visually impaired using computer vision." 2020 4th International Conference on Trends in Electronics and Informatics (ICOEI) (48184). IEEE, 2020.
3. Chaganti, Sai Yeshwanth, et al. "Image Classification using SVM and CNN." 2020 International conference on computer science, engineering, and applications (ICCSEA). IEEE, 2020.
4. Bhardwaj, Deepika, and Vinod Todwal. "A Novel Deep Learning Model for Indoor-Outdoor Scene Classification Using VGG-16 Deep CNN." IJRTE, 2021.
5. Simonyan, Karen, and Andrew Zisserman. "Very deep convolutional networks for large-scale image recognition." arXiv preprint arXiv:1409.1556 (2014).
6. Mikołajczyk, A.; Grochowski, M. Data Augmentation for Improving Deep Learning in Image Classification Problem. In Proceedings of the International Interdisciplinary PhD Workshop (IIPhDW), Swinoujście, Poland, 9–12 May 2018; pp. 117–122.
7. Narayan Gupta, Shesh, and Nicholas Bear Brown. "Adjusting for Bias with Procedural Data." arXiv e-prints (2022): arXiv-2204.
8. A. Quattoni, and A.Torralba. Recognizing Indoor Scenes. IEEE Conference on Computer Vision and Pattern Recognition (CVPR), 2009.
9. Zhou, Bolei, et al. "Places: A 10 million image database for scene recognition." IEEE transactions on pattern analysis and machine intelligence 40.6 (2017): 1452-1464.